\documentclass{article}%
\usepackage{amsmath}
\usepackage{amsfonts}
\usepackage{amssymb}
\usepackage{graphicx}%
\providecommand{\U}[1]{\protect\rule{.1in}{.1in}}
\newtheorem{theorem}{Theorem} [section]

\newtheorem{corollary}[theorem]{Corollary}

\newtheorem{lemma}[theorem]{Lemma}

\newtheorem{proposition}[theorem]{Proposition}

\newenvironment{proof}[1][Proof]{\noindent\textbf{#1.} }{\ \rule{0.5em}{0.5em}}
\begin{document}

\title{On the independence polynomial of an antiregular graph}
\author{Vadim E. Levit\\Ariel University Center of Samaria, Israel\\levitv@ariel.ac.il
\and Eugen Mandrescu\\Holon Institute of Technology, Israel\\eugen\_m@hit.ac.il}
\date{}
\maketitle

\begin{abstract}
A graph with at most two vertices of the same degree is called
\textit{antiregular} \cite{Merris2003}, \textit{maximally nonregular}
\cite{Zykov} or \textit{quasiperfect} \cite{Behzad}. If $s_{k}$ is the number
of independent sets of cardinality $k$ in a graph $G$, then
\[
I(G;x)=s_{0}+s_{1}x+...+s_{\alpha}x^{\alpha}%
\]
is the \textit{independence polynomial} of $G$ \cite{GutHar83}, where
$\alpha=\alpha(G)$ is the size of a maximum independent set.

In this paper we derive closed formulae for the independence polynomials of
antiregular graphs. In particular, we deduce that every antiregular graph $A$
is uniquely defined by its independence polynomial $I(A;x)$, within the family
of threshold graphs. Moreover, $I(A;x)$ is log-concave with at most two real
roots, and $I(A;-1)\in\{-1,0\}$.


\end{abstract}

\section{Introduction}

Throughout this paper $G=(V,E)$ is a simple (i.e., a finite, undirected,
loopless and without multiple edges) graph with vertex set $V=V(G)$ and edge
set $E=E(G)$. The \textit{neighborhood} of a vertex $v\in V$ is the set
$N_{G}(v)=\{w:w\in V$ \ \textit{and} $vw\in E\}$, and $N_{G}[v]=N_{G}%
(v)\cup\{v\}$; if there is ambiguity on $G$, we use $N(v)$ and $N[v]$,
respectively. If $\left\vert N(v)\right\vert =1$, then $v$ is a
\textit{pendant vertex} of $G$ and $v$ is a \textit{simplicial vertex} if
$G[N[v]]$ is a complete graph. A maximal clique containing at least simplicial
vertex is called a \textit{simplex}. By \textrm{simp}$(G)$ we mean the set of
all simplicial vertices of $G$. A graph $G$ is said to be \textit{simplicial}
if every vertex of $G$ is simplicial or is adjacent to a simplicial vertex
\cite{CheHaHeLas}.

$\overline{G}$ stands for the complement of $G$. $K_{n}$, $K_{m,n}$, $P_{n}$
denote the complete graph on $n\geq1$ vertices, the complete bipartite graph
on $m,n\geq1$ vertices, and the chordless path on $n\geq1$ vertices, respectively.

The \textit{disjoint union} of the graphs $G_{1},G_{2}$ is the graph
$G=G_{1}\cup G_{2}$ having the disjoint union of $V(G_{1}),V(G_{2})$ as a
vertex set, and the disjoint union of $E(G_{1}),E(G_{2})$ as an edge set. In
particular, $nG$ denotes the disjoint union of $n>1$ copies of the graph $G$. 

If $G_{1},G_{2}$ are disjoint graphs, then their \textit{Zykov sum} is the
graph $G_{1}+G_{2}$ with $V(G_{1})\cup V(G_{2})$ as a vertex set and
$E(G_{1})\cup E(G_{2})\cup\{v_{1}v_{2}:v_{1}\in V(G_{1}),v_{2}\in V(G_{2})\}$
as an edge set.

A set of pairwise non-adjacent vertices is called \textit{independent}. If $S$
is an independent set, then we denote $N(S)=\{v:N(v)\cap S\neq\emptyset\}$ and
$N[S]=N(S)\cup S$.

An independent set of maximum size will be referred to as a \textit{maximum
independent set} of $G$. The \textit{independence number }of $G$, denoted by
$\alpha(G)$, is the number of vertices of a maximum independent set in $G$.

A \textit{matching} is a set of non-incident edges of $G$, and $\mu(G)$ is the
maximum cardinality of a matching. $G$ is a \textit{K\"{o}nig-Egerv\'{a}ry
graph }provided $\alpha(G)+\mu(G)=\left\vert V(G)\right\vert $, \cite{dem},
\cite{ster}.

Let $s_{k}$ be the number of independent sets in $G$ of cardinality
$k\in\{0,1,...,\alpha(G)\}$. The polynomial
\[
I(G;x)=s_{0}+s_{1}x+s_{2}x^{2}+...+s_{\alpha}x^{\alpha},\alpha=\alpha(G),
\]
is called the \textit{independence polynomial} of $G$ (Gutman and Harary
\cite{GutHar83}).

For a survey on independence polynomials the reader is referred to
\cite{LeMa05}. \begin{figure}[h]
\setlength{\unitlength}{1cm}\begin{picture}(5,1.8)\thicklines
\multiput(3,0.5)(1,0){4}{\circle*{0.29}}
\multiput(4,1.5)(1,0){2}{\circle*{0.29}}
\put(3,0.5){\line(1,0){3}} \put(4,0.5){\line(0,1){1}}
\put(4,1.5){\line(1,0){1}} \put(5,0.5){\line(0,1){1}}
\put(3.75,1){\makebox(0,0){$c$}} \put(5.2,1){\makebox(0,0){$b$}}
\put(3.5,0.2){\makebox(0,0){$d$}}
\put(4.5,1.3){\makebox(0,0){$a$}}
\put(5.5,0.2){\makebox(0,0){$f$}}
\put(4.5,0.2){\makebox(0,0){$e$}}
\put(2.3,1){\makebox(0,0){$G_{1}$}}
\multiput(8,0.5)(1,0){4}{\circle*{0.29}}
\multiput(9,1.5)(1,0){2}{\circle*{0.29}}
\put(8,0.5){\line(1,1){1}} \put(8,0.5){\line(2,1){2}}
\put(9,0.5){\line(1,0){2}} \put(9,0.5){\line(0,1){1}}
\put(9,1.5){\line(1,0){1}} \put(10,0.5){\line(0,1){1}}
\put(10,1.5){\line(1,-1){1}} \put(8,0.1){\makebox(0,0){$f$}}
\put(8.7,1.5){\makebox(0,0){$b$}} \put(10,0.1){\makebox(0,0){$c$}}
\put(11,0.1){\makebox(0,0){$d$}} \put(9,0.1){\makebox(0,0){$a$}}
\put(10.3,1.5){\makebox(0,0){$e$}}
\put(7.3,1){\makebox(0,0){$G_{2}$}}
\end{picture}
\caption{$G_{2}$ is the line-graph of and $G_{1}$.}%
\label{fig1}%
\end{figure}
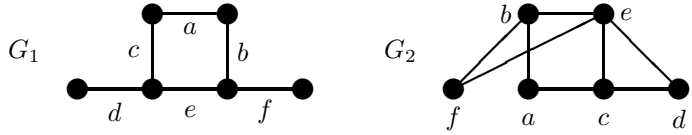

The independence polynomial is a generalization of the matching polynomial
\cite{GutHar83}, because the matching polynomial of a graph and the
independence polynomial of its line graph are identical. Recall that given a
graph $G$, its \textit{line graph} $L(G)$ is the graph whose vertex set is the
edge set of $G$, and two vertices are adjacent if they share an end vertex in
$G$. For instance, the graphs $G_{1}$ and $G_{2}$ depicted in Figure
\ref{fig1} satisfy $G_{2}=L(G_{1})$ and, hence,
\[
I(G_{2};x)=1+6x+7x^{2}+x^{3}=M(G_{1};x),
\]
where $M(G_{1};x)$ is the matching polynomial of the graph $G_{1}$. In
\cite{GutHar83} a\emph{\ }number of general properties of the independence
polynomial of a graph are presented. As examples, we mention that:%
\[
I(G_{1}\cup G_{2};x)=I(G_{1};x)\cdot I(G_{2};x),\quad I(G_{1}+G_{2}%
;x)=I(G_{1};x)+I(G_{2};x)-1.
\]
The following equality, due to Gutman and Harary \cite{GutHar83}, is very
useful in calculating the independence polynomial for various families of graphs.

\begin{proposition}
\cite{GutHar83}\label{prop1} If $w\in V(G)$, then $I(G;x)=I(G-w;x)+x\cdot
I(G-N[w];x)$.
\end{proposition}

A finite sequence of real numbers $(a_{0},a_{1},a_{2},...,a_{n})$ is said to be:

\begin{itemize}
\item \textit{unimodal} if there is some $k\in\{0,1,...,n\}$, called the
\textit{mode} of the sequence, such that
\[
a_{0}\leq...\leq a_{k-1}\leq a_{k}\geq a_{k+1}\geq...\geq a_{n};
\]

\item \textit{log-concave} if $a_{i}^{2}\geq a_{i-1}\cdot a_{i+1}$ for
$i\in\{1,2,...,n-1\}$.
\end{itemize}

It is known that every log-concave sequence of positive numbers is unimodal as well.

A polynomial is called \textit{unimodal (log-concave)} if the sequence of its
coefficients is unimodal (log-concave, respectively). The product of two
unimodal polynomials is not always unimodal, even if they are independence
polynomials; e.g.,
\[
I(K_{100}+3K_{7};x)=1+121x+147x^{2}+343x^{3}%
\]
and
\[
I(K_{120}+3K_{7};x)=1+141x+147x^{2}+343x^{3},
\]
while their product is not unimodal:%

\[
1+262x+17\,355x^{2}+39\,200\allowbreak x^{3}+111\,475x^{4}+\mathbf{100\,842}%
x^{5}+117\,649x^{6}.
\]

\begin{theorem}
\cite{KeilsonGerber}\label{th1} If $P,Q$ are polynomials, such that $P$ is
log-concave and $Q$ is unimodal, then $P\cdot Q$ is unimodal, while the
product of two log-concave polynomials is log-concave.
\end{theorem}

Alavi, Malde, Schwenk and Erd\"{o}s \cite{AlMalSchErdos} proved that for any
permutation $\pi$ of $\{1,2,...,\alpha\}$ there is a graph $G$ with
$\alpha(G)=\alpha$, such that the coefficients of $I(G;x)$ satisfy
\[
s_{\pi(1)}<s_{\pi(2)}<s_{\pi(3)}<...<s_{\pi(\alpha)}.
\]
For instance, the independence polynomial:

\begin{itemize}
\item $I(K_{42}+3K_{7};x)=1+63x+147x^{2}+343x^{3}$ is log-concave;

\item $I(K_{43}+3K_{7};x)=1+64x+147x^{2}+343x^{3}$ is unimodal, but
non-log-concave, because $147\cdot147-64\cdot343=-343<0$;

\item $I(K_{127}+3K_{7};x)=1+148x+147x^{2}+343x^{3}$ is non-unimodal.
\end{itemize}

Some more discussion on independence polynomials may be found in
\cite{BrownDilNow}, \cite{LeMa02}, \cite{LeMa03a}, \cite{LeMa03b},
\cite{LeMa04}, \cite{LeMa06a}, \cite{LeMa06b}, \cite{LeMa06c}, \cite{LeMa07},
\cite{LeMa08a}, and \cite{LeMa08b}.

\begin{figure}[h]
\setlength{\unitlength}{1cm}\begin{picture}(5,2)\thicklines
\multiput(3,0.5)(1,0){2}{\circle*{0.29}}
\multiput(2,1.5)(1,0){3}{\circle*{0.29}}
\put(3,0.5){\line(1,0){1}}
\put(3,0.5){\line(-1,1){1}}
\put(3,0.5){\line(1,1){1}}
\put(3,0.5){\line(1,1){1}}
\put(3,0.5){\line(0,1){1}}
\put(4,0.5){\line(-2,1){2}}
\put(4,0.5){\line(-1,1){1}}
\put(4,0.5){\line(0,1){1}}
\put(3,0){\makebox(0,0){$v_{4}$}}
\put(4,0){\makebox(0,0){$v_{5}$}}
\put(2,1.9){\makebox(0,0){$v_{1}$}}
\put(3,1.9){\makebox(0,0){$v_{2}$}}
\put(4,1.9){\makebox(0,0){$v_{3}$}}
\put(1.3,1){\makebox(0,0){$G_{1}$}}
\multiput(6,0.5)(1,0){3}{\circle*{0.29}}
\multiput(6,1.5)(1,0){2}{\circle*{0.29}}
\put(6,0.5){\line(1,1){1}}
\put(6,0.5){\line(0,1){1}}
\put(6,0.5){\line(1,0){1}}
\put(6,1.5){\line(1,0){1}}
\put(7,0.5){\line(-1,1){1}}
\put(7,0.5){\line(0,1){1}}
\put(7,1.5){\line(1,-1){1}}
\put(6,0.1){\makebox(0,0){$v_{2}$}}
\put(7,0.1){\makebox(0,0){$v_{3}$}}
\put(8,0.1){\makebox(0,0){$v_{4}$}}
\put(6,1.9){\makebox(0,0){$v_{1}$}}
\put(7,1.9){\makebox(0,0){$v_{5}$}}
\put(5.3,1){\makebox(0,0){$G_{2}$}}
\multiput(10,0.5)(1,0){3}{\circle*{0.29}}
\multiput(10,1.5)(1,0){2}{\circle*{0.29}}
\put(10,0.5){\line(1,1){1}}
\put(10,0.5){\line(0,1){1}}
\put(10,0.5){\line(1,0){2}}
\put(10,1.5){\line(1,0){1}}
\put(11,0.5){\line(-1,1){1}}
\put(11,0.5){\line(0,1){1}}
\put(11,1.5){\line(1,-1){1}}
\put(10,0.1){\makebox(0,0){$v_{2}$}}
\put(11,0.1){\makebox(0,0){$v_{4}$}}
\put(12,0.1){\makebox(0,0){$v_{3}$}}
\put(10,1.9){\makebox(0,0){$v_{1}$}}
\put(11,1.9){\makebox(0,0){$v_{5}$}}
\put(9.3,1){\makebox(0,0){$G_{3}$}}
\end{picture}\caption{$G_{1},G_{2}$ and $G_{3}$ are threshold graphs
corresponding to the binary building strings $00011,01101,01011$,
respectively.}%
\label{fig111}%
\end{figure}
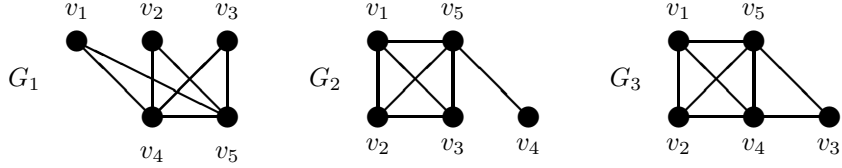

Recall that $G$ is a \textit{threshold} graph if it can be obtained from
$K_{1}$ by iterating the operations of complementation and disjoint union with
a new copy of $K_{1}$ in any order \cite{ChvatHam}. In other words, $G$ is a
threshold graph if it can be obtained from $K_{1}$ by iterating the operations
of adding in a new vertex which is connected to no other vertex (i.e., an
\textit{isolated vertex}) or adding in a new vertex connected to every other
vertex (i.e., a \textit{cone vertex}, or a \textit{universal vertex}, or a
\textit{dominating vertex}). The sequence of operations which describes this
process can be represented as a binary string, which we call the
\textit{binary building string} of $G$, where "$0$" means "\textit{adding an
isolated vertex}" and "$1$" corresponds to "\textit{adding a dominating
vertex}". Clearly, each such a string begins by a "$0$". For some examples see
Figure \ref{fig111}, where\emph{ }the vertex $v_{i}$ is dominating if and only
if the $i^{th}$ bit in the binary building string equals "$1$". Chvatal and
Hammer \cite{ChvatHam} showed that threshold graphs are exactly the graphs
having no induced subgraph isomorphic to either a $P_{4}$, or a $C_{4}$, or a
$2K_{2}$.

Following Hoede and Li \cite{HoLi94}, $G$ is called a \textit{clique-unique
graph} if the relation $I(\overline{G};x)=I(\overline{H};x)$ implies that
$\overline{G}$ and $\overline{H}$ are isomorphic (or, equivalently, $G$ and
$H$ are isomorphic). One of the problems they proposed was to determine
clique-unique graphs (Problem 4.1, \cite{HoLi94}). In \cite{LeMa08b} it is
proved that spiders are \textit{independence-unique graphs }within the family
of well-covered trees,\textit{ }i.e., they are uniquely defined by their
independence polynomials in the context of well-covered trees. The following
result, due to Stevanovi\'{c}, says that every threshold graph is completely
determined by its independence polynomial within the class of threshold graphs.

\begin{theorem}
\label{th5}\cite{Stevan} Two threshold graphs have the same independence
polynomial if and only if they are isomorphic.
\end{theorem}

It is well-known that every graph of order at least two has at least two
vertices of the same degree. A graph having at most two vertices of the same
degree is called \textit{antiregular} \cite{Merris2001}, \cite{Merris2003},
\textit{maximally nonregular} \cite{Zykov} or \textit{quasiperfect}
\cite{Behzad}, \cite{Nebesky}, \cite{Sedlacek}. Some examples of antiregular
graphs are presented in Figure \ref{Fig12}. 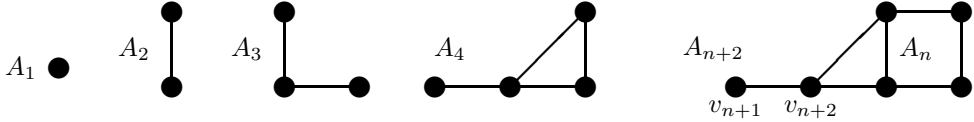
\begin{figure}[h]
\setlength{\unitlength}{1cm}\begin{picture}(5,1.8)\thicklines
\put(1,0.75){\circle*{0.29}}
\put(0.5,0.75){\makebox(0,0){$A_{1}$}}
\multiput(2.5,0.5)(0,1){2}{\circle*{0.29}}
\put(2.5,0.5){\line(0,1){1}}
\put(2,1){\makebox(0,0){$A_{2}$}}
\multiput(4,0.5)(0,1){2}{\circle*{0.29}}
\put(5,0.5){\circle*{0.29}}
\put(4,0.5){\line(0,1){1}}
\put(4,0.5){\line(1,0){1}}
\put(3.5,1){\makebox(0,0){$A_{3}$}}
\multiput(6,0.5)(1,0){3}{\circle*{0.29}}
\put(8,1.5){\circle*{0.29}}
\put(6,0.5){\line(1,0){2}}
\put(7,0.5){\line(1,1){1}}
\put(8,0.5){\line(0,1){1}}
\put(6.2,1){\makebox(0,0){$A_{4}$}}
\multiput(10,0.5)(1,0){4}{\circle*{0.29}}
\put(12,1.5){\circle*{0.29}}
\put(13,1.5){\circle*{0.29}}
\put(10,0.5){\line(1,0){3}}
\put(11,0.5){\line(1,1){1}}
\put(12,0.5){\line(0,1){1}}
\put(12,1.5){\line(1,0){1}}
\put(13,0.5){\line(0,1){1}}
\put(10,0.2){\makebox(0,0){$v_{n+1}$}}
\put(11,0.2){\makebox(0,0){$v_{n+2}$}}
\put(12.4,1){\makebox(0,0){$A_{n}$}}
\put(9.7,1){\makebox(0,0){$A_{n+2}$}}
\end{picture}\caption{Antiregular graphs: $A_{1},A_{2},A_{3},A_{4}$ and
$A_{n+2}=K_{1}+(K_{1}\cup A_{n})$.}%
\label{Fig12}%
\end{figure}

It is intuitively clear that the number of different antiregular graphs of the
same order is quite small.

\begin{theorem}
\cite{Behzad} For every positive integer $n\geq2$ there is a unique connected
antiregular graph of order $n$, denoted by $A_{n}$, and a unique non-connected
antiregular graph of order $n$, namely $\overline{A_{n}}$.
\end{theorem}

Moreover, antiregular graphs enjoy a very specific recursive structure.

\begin{theorem}
\label{th2}\cite{Merris2003} The antiregular graphs can be defined by the
following recurrences:%
\begin{align*}
A_{1} &  =K_{1},A_{n+1}=K_{1}+\overline{A_{n}},n\geq1\text{, or }\\
A_{1} &  =K_{1},A_{2}=K_{2},A_{n+2}=K_{1}+(K_{1}\cup A_{n}),n\geq1.
\end{align*}

\end{theorem}

Characteristic, admittance and matching polynomials of antiregular graphs were
studied in \cite{Munarini09}.

In this paper we present closed formulae for the independence polynomial of an
antiregular graph. We also show that independence polynomials of antiregular
graphs are log-concave. Moreover, it turns out that antiregular graphs are
completely determined by their independence polynomials within the class of
threshold graphs.

\section{Results}

Antiregular graphs have a number of nice properties. Some of them are
presented in \cite{Behzad}, \cite{Merris2001}, \cite{Merris2003},
\cite{Nebesky}, \cite{Sedlacek}.

\begin{theorem}
\label{th3}Every antiregular graph is threshold, simplicial, and a
K\"{o}nig-Egerv\'{a}ry graph.
\end{theorem}

\begin{proof}
\begin{itemize}
\item The building binary strings of the forms $\mathbf{00}1010101...$ and
$\mathbf{01}010101...$ give rise to antiregular graphs. Hence, any antiregular
graph is threshold.

\item We prove that every antiregular graph is simplicial, by induction on
$n$. Clearly, $A_{1}=K_{1}$, $A_{2}=K_{2}$ are simplicial graphs. Since, by
Theorem \ref{th2}, we have $A_{3}=K_{1}+(K_{1}\cup K_{1})$, it is easy to see
that $A_{3}=P_{3}$ and, consequently, $A_{3}$ is a simplicial graph.

Assume that the assertion is true for $2\leq m\leq n+1$.

Since $A_{n+2}=K_{1}+(K_{1}\cup A_{n}),n\geq1$, it follows that $A_{n+2}$ has
a vertex of degree one, say $v_{n+2}$ and a vertex of degree $n+1$, say
$v_{n+1}$. Clearly, $v_{n+2}$ is a simplicial vertex, as it is a pendant
vertex, while $v_{n+1}$ is not a simplicial one. On the other hand, if $v\in$
\textrm{simp}$(A_{n})$, then $N_{A_{n}}[v]$ induces a complete subgraph of
$A_{n}$. Hence, we infer that
\[
N_{A_{n+2}}\left[  v\right]  =N_{A_{n}}\left[  v\right]  \cup\{v_{n+1}\}
\]
induces a complete subgraph of $A_{n+2}$, since $N_{A_{n+2}}\left[
v_{n+1}\right]  =V(A_{n+2})-\{v_{n+1}\}$. Consequently, it follows that
\[
\mathrm{simp}(A_{n+2})=\mathrm{simp}(A_{n})\cup\{v_{n+2}\}.
\]
In other words, each vertex of $A_{n+2}$ is either a simplicial vertex or it
is adjacent to a simplicial vertex. Therefore, $A_{n}$ is a simplicial graph,
for every $n\geq1$. Since, clearly, $\overline{A_{n}}=\overline{K_{1}%
+\overline{A_{n-1}}}=\overline{K_{1}}\cup\overline{\overline{A_{n-1}}}%
=K_{1}\cup A_{n-1}$, we get that $\overline{A_{n}},n\geq1$, is a simplicial
graph, too.

\item It is easy to see that the independence number $\alpha\left(
A_{n}\right)  $ is equal to $\left\lceil \frac{n}{2}\right\rceil $. On the
other hand, using the fact that $A_{n}=K_{1}+(K_{1}\cup A_{n-2}),n\geq3$, one
can easily see that its matching number $\mu\left(  A_{n}\right)  $ equals
$\left\lfloor \frac{n}{2}\right\rfloor $. Since $\left\lceil \frac{n}%
{2}\right\rceil +\left\lfloor \frac{n}{2}\right\rfloor =n$, the graph $A_{n}$
is a\ K\"{o}nig-Egerv\'{a}ry graph.
\end{itemize}
\end{proof}

Using the recursive structure of antiregular graphs we get the following.

\begin{lemma}
\label{lemma1}$I(A_{n+2};x)=\left(  1+x\right)  \bullet\left(  1+I(A_{n}%
;x)\right)  -1$ holds for every $n\geq1$.
\end{lemma}

\begin{proof}
Clearly, $I(A_{1};x)=I(K_{1};x)=1+x$, while $I(A_{2};x)=I(K_{2};x)=1+2x$.
Further, according to Proposition \ref{prop1} and Theorem \ref{th2}, we infer
that:%
\begin{align*}
I(A_{n+2};x) &  =I(K_{1}+(K_{1}\cup A_{n});x)=I(K_{1};x)+I(K_{1}\cup
A_{n};x)-1=\\
&  =I(K_{1};x)+I(K_{1};x)\bullet I(A_{n};x)-1=\left(  1+x\right)
\bullet\left(  1+I(A_{n};x)\right)  -1,
\end{align*}
as required.
\end{proof}

Using Lemma \ref{lemma1}, one can easily compute the following independence
polynomials:%
\begin{align*}
I(A_{3};x) &  =1+3x+x^{2},\qquad\qquad\qquad\quad I(A_{4};x)=1+4x+2x^{2}\\
I(A_{5};x) &  =1+5x+4x^{2}+x^{3},\qquad\quad\quad I(A_{6};x)=1+6x+6x^{2}%
+2x^{3}\\
I(A_{7};x) &  =1+7x+9x^{2}+5x^{3}+x^{4},\quad I(A_{8};x)=1+8x+12x^{2}%
+8x^{3}+2x^{4}.
\end{align*}

\begin{theorem}
\label{th4}The independence polynomial of $A_{n}$ is:%
\begin{align*}
I(A_{2k-1};x)  &  =\left(  1+x\right)  ^{k}+\left(  1+x\right)  ^{k-1}%
-1,k\geq1,\\
I(A_{2k};x)  &  =2\cdot\left(  1+x\right)  ^{k}-1,k\geq1,
\end{align*}
The independence polynomial of $\overline{A_{n}}$ is
\[
I(\overline{A_{n}};x)=\left(  1+x\right)  \cdot I(A_{n-1};x),n\geq2,
\]
and, consequently,
\begin{align*}
I(\overline{A_{2k-1}};x)  &  =2\cdot\left(  1+x\right)  ^{k}-x-1,k\geq1,\\
I(\overline{A_{2k}};x)  &  =\left(  1+x\right)  ^{k+1}+\left(  1+x\right)
^{k}-x-1,k\geq1.
\end{align*}

\end{theorem}

\begin{proof}
We prove the formulae for $I(A_{n};x)$ by induction on $n$.

If $n\in\{1,2\}$, then
\[
I(A_{1};x)=I(K_{1};x)=1+x=\left(  1+x\right)  ^{1}+\left(  1+x\right)  ^{0}-1
\]
and
\[
I(A_{2};x)=I(K_{2};x)=1+2x=2\cdot\left(  1+x\right)  ^{1}-1.
\]

Assume that the formulae are true for $2\leq$ $m\leq n$.

Let $n+1=2k+1$. Then, by Lemma \ref{lemma1} and induction hypothesis, we
obtain:
\begin{align*}
I(A_{n+1};x)  &  =I(A_{2k+1};x)=\left(  1+x\right)  \bullet\left(
1+I(A_{2k-1};x)\right)  -1=\\
&  =\left(  1+x\right)  \cdot(1+\left(  1+x\right)  ^{k}+\left(  1+x\right)
^{k-1}-1)-1=\\
&  =\left(  1+x\right)  ^{k+1}+\left(  1+x\right)  ^{k}-1.
\end{align*}

Let $n+1=2k$. Then again, using Lemma \ref{lemma1} and the induction
hypothesis, we get:
\begin{align*}
I(A_{n+1};x)  &  =I(A_{2k};x)=\left(  1+x\right)  \bullet\left(
1+I(A_{2k-2};x)\right)  -1=\\
&  =\left(  1+x\right)  \cdot(1+2\cdot\left(  1+x\right)  ^{k-1}-1)-1=\\
&  =2\cdot\left(  1+x\right)  ^{k}-1.
\end{align*}

In conclusion, both formulae are true.

According to Theorem \ref{th2}, we have $A_{n}=K_{1}+\overline{A_{n-1}}$,
which implies
\[
\overline{A_{n}}=\overline{K_{1}+\overline{A_{n-1}}}=\overline{K_{1}}%
\cup\overline{\overline{A_{n-1}}}=K_{1}\cup A_{n-1},
\]
and hence, we deduce that $I(\overline{A_{n}};x)=(1+x)\cdot I(A_{n-1};x)$.
Further, using the result obtained for $I(A_{n-1};x)$, we finally infer the
closed formulae for $I(\overline{A_{n}};x)$, as claimed.
\end{proof}

The \textit{Fibonacci number} of a graph $G$ is the number of all its
independent sets \cite{ProdTichy}. Obviously, the Fibonacci number of $G$ is
equal to
\[
I(G;1)=s_{0}+s_{1}+s_{2}+...+s_{\alpha(G)},
\]
where
\[
I(G;x)=s_{0}+s_{1}x+s_{2}x^{2}+...+s_{\alpha(G)}x^{\alpha(G)}%
\]
is the independence polynomial of $G$. Using Theorem \ref{th4}, we immediately
obtain the following.

\begin{corollary}
The Fibonacci numbers of $A_{n}$ are:%
\[
I(A_{2k-1};1)=3\bullet2^{k-1}-1\text{ and }I(A_{2k};1)=2^{k+1}-1,k\geq1,
\]
while the Fibonacci numbers of $\overline{A_{n}}$ are:%
\[
I(\overline{A_{2k-1}};1)=2^{k+1}-2\text{ and }I(\overline{A_{2k}}%
;1)=3\bullet2^{k}-2,k\geq1.
\]

\end{corollary}

If $G$ has $s_{k}$ independent sets of size $k$, then
\[
s_{0}-s_{1}+s_{2}-s_{3}+...+(-1)^{\alpha(G)}s_{\alpha(G)}%
\]
is called the \textit{alternating number of independent sets} of $G$
\cite{BSN07}. Evidently,
\begin{align*}
I(G;-1) &  =s_{0}-s_{1}+s_{2}-s_{3}+s_{4}-...+\left(  -1\right)  ^{\alpha
(G)}s_{\alpha(G)}\\
&  =(s_{0}+s_{2}+s_{4}+...)-\left(  s_{1}+s_{3}+s_{5}+...\right)  .
\end{align*}
In addition, if we denote
\begin{align*}
\mathrm{even}(G)  & =s_{0}+s_{2}+s_{4}+...,\\
\mathrm{odd}(G)  & =s_{1}+s_{3}+s_{5}+...,
\end{align*}
then we may conclude that:
\[
\text{"\textit{the alternating number of independent sets of} }G\text{"}%
=I(G;-1)=\mathrm{even}(G)-\mathrm{odd}(G).
\]

Let us notice that the difference $\left\vert \mathrm{even}(G)-\mathrm{odd}%
(G)\right\vert $ can be indefinitely large. For instance, $I(K_{n};x)=1+nx$
and hence, $I(K_{n};-1)=1-n\leq0$. On the other hand, the graph $H=\left(
K_{m}\cup K_{n}\right)  +K_{1}$ has
\[
I(H;x)=\left(  1+mx\right)  \left(  1+nx\right)  +x
\]
and, consequently,%
\[
I(H;-1)=\left(  1-m\right)  \left(  1-n\right)  -1>0,m>2,n>1.
\]

\begin{corollary}
For a connected antiregular graph the number of independent sets of odd size
is greater by one than the number of independent sets of even size, while for
a disconnected antiregular graph, the two numbers are equal.
\end{corollary}

\begin{proof}
Let denote by $\alpha$ the independence number of $A_{n}$, i.e.,
$\alpha=\alpha(A_{n})=\left\lceil n/2\right\rceil $, and $I(A_{n}%
;x)=s_{0}+s_{1}x+s_{2}x^{2}+...+s_{\alpha}x^{\alpha}$.

According to Theorem \ref{th4}, it follows that $I(A_{n};-1)=-1$, which
clearly implies
\[
(s_{0}+s_{2}+s_{4}+...)+1=s_{1}+s_{3}+s_{5}+...,
\]
i.e., \textrm{even}$(G)+1=$ \textrm{odd}$(G)$, as required.

Similarly, in accordance with Theorem \ref{th4}, we obtain that $I(\overline
{A_{n}};-1)=0$, which ensures that
\[
s_{0}+s_{2}+s_{4}+...=s_{1}+s_{3}+s_{5}+...,
\]
i.e., \textrm{even}$(G)=$ \textrm{odd}$(G)$, and this completes the proof.
\end{proof}

Let us notice that there are threshold graphs, whose independence polynomials are

\begin{itemize}
\item non-unimodal, e.g., $G=6K_{1}+K_{10}$, whose binary building string is
$6[0]10[1]$ and independence polynomial is
\[
I(G;x)=\left(  1+x\right)  ^{6}+10x=x^{6}+6x^{5}+15x^{4}+20x^{3}%
+\mathbf{15}x^{2}+16x+1;
\]

\item unimodal, but non-log-concave, e.g., $G=3K_{1}+K_{7}$, whose binary
building string is $3[0]7[1]$ and independence polynomial is%
\[
I(G;x)=\left(  1+x\right)  ^{3}+7x=x^{3}+\mathbf{3}x^{2}+10x+1;
\]

\item log-concave, e.g., $G=7K_{1}+K_{5}$, whose binary building string is
$7[0]5[1]$ and independence polynomial is%
\[
I(G;x)=\left(  1+x\right)  ^{7}+5x=x^{7}+7x^{6}+21x^{5}+35x^{4}+35x^{3}%
+21x^{2}+12x+1.
\]

\end{itemize}

\begin{corollary}
The independence polynomials of $A_{n}$ and $\overline{A_{n}}$ are
log-concave, for every integer $n\geq1$.
\end{corollary}

\begin{proof}
According to Theorem \ref{th4}, $I(A_{2k};x)=2\cdot\left(  1+x\right)  ^{k}-1$
and hence, $I(A_{2k-1};x)$ is log-concave, because $\left(  1+x\right)  ^{k}$
is log-concave. The polynomial
\[
I(A_{2k-1};x)=\left(  1+x\right)  ^{k}+\left(  1+x\right)  ^{k-1}-1=\left(
1+x\right)  ^{k-1}\left(  2+x\right)  -1
\]
is log-concave, since the product of two log-concave polynomials is again
log-concave, by Theorem \ref{th1}.

Similarly, $I(\overline{A_{n}};x)$ is log-concave, because $I(\overline{A_{n}%
};x)=\left(  1+x\right)  \cdot I(A_{n-1};x)$.
\end{proof}

\begin{figure}[h]
\setlength{\unitlength}{1cm}\begin{picture}(5,1.2)\thicklines
\multiput(1,0)(1,0){6}{\circle*{0.29}}
\multiput(1,1)(2,0){2}{\circle*{0.29}}
\multiput(4,1)(2,0){2}{\circle*{0.29}}
\put(1,0){\line(1,0){5}}
\put(1,0){\line(0,1){1}}
\put(3,0){\line(0,1){1}}
\put(4,0){\line(0,1){1}}
\put(6,0){\line(0,1){1}}
\put(0.5,0.3){\makebox(0,0){$T_{1}$}}
\multiput(7.6,0)(1,0){6}{\circle*{0.29}}
\multiput(9.6,1)(1,0){3}{\circle*{0.29}}
\put(7.6,1){\circle*{0.29}}
\put(7.6,0){\line(1,0){5}}
\put(7.6,0){\line(0,1){1}}
\put(9.6,1){\line(1,0){1}}
\put(10.6,0){\line(0,1){1}}
\put(11.6,0){\line(0,1){1}}
\put(7.1,0.3){\makebox(0,0){$T_{2}$}}
\end{picture}
\caption{Non-isomorphic trees with the same independence polynomial.}%
\label{fig505}%
\end{figure}
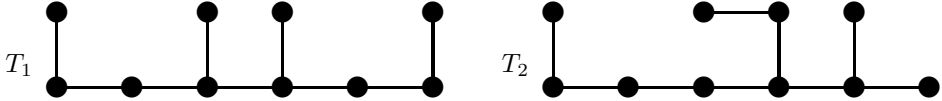

Let us mention that there are non-isomorphic graphs with the same independence
polynomial. For instance, Dohmen, P\"{o}nitz and Tittmann \cite{DoPoTi03} have
found two non-isomorphic trees (Figure \ref{fig505}) having the same
independence polynomial, namely,
\[
I(T_{1};x)=I(T_{2};x)=1+10x+36x^{2}+58x^{3}+42x^{4}+12x^{5}+x^{6}.
\]

Let us notice that $I(A_{2k};x)=I(K_{k,k};x)$ and $I(A_{2k-1};x)=I(K_{k,k-1}%
;x)$. For $k\geq3$ neither $K_{k,k}$ nor $K_{k,k-1}$ is a threshold graph,
because they contain an induced subgraph isomorphic to $C_{4}$.

\begin{figure}[h]
\setlength{\unitlength}{1cm}\begin{picture}(5,1.2)\thicklines
\multiput(4,0)(1,0){3}{\circle*{0.29}}
\multiput(5,1)(1,0){2}{\circle*{0.29}}
\put(4,0){\line(2,1){2}}
\put(4,0){\line(1,1){1}}
\put(5,0){\line(0,1){1}}
\put(5,0){\line(1,1){1}}
\put(5,1){\line(1,-1){1}}
\put(6,0){\line(0,1){1}}
\put(3.2,0.5){\makebox(0,0){$K_{2,3}$}}
\multiput(9,0)(1,0){3}{\circle*{0.29}}
\multiput(10,1)(1,0){2}{\circle*{0.29}}
\put(9,0){\line(1,0){2}}
\put(10,0){\line(0,1){1}}
\put(10,0){\line(1,1){1}}
\put(10,1){\line(1,0){1}}
\put(10,1){\line(1,-1){1}}
\put(8.2,0.5){\makebox(0,0){$A_{5}$}}
\end{picture}\caption{Non-isomorphic graphs having $I(K_{2,3};x)=I(A_{5}%
;x)=1+5x+4x^{2}+x^{3}$.}%
\label{fig22}%
\end{figure}
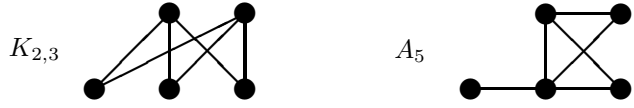

Using Theorems \ref{th5} and \ref{th3}, we infer the following.

\begin{corollary}
Every antiregular graph is a unique-independence graph within the family of
threshold graphs, i.e., if a threshold graph $G$ has $I(G;x)=I(A_{n};x)$ or
$I(G;x)=I(\overline{A_{n}};x)$, then $G$ is isomorphic to $A_{n}$ or
$\overline{A_{n}}$, respectively.
\end{corollary}

It is known that the independence polynomial has at least one real root
\cite{FishSol90}.

\begin{corollary}
\label{Cor1}The polynomial $I(A_{2k};x)$ has only one real root for every odd
$k$ and exactly two real roots for each even $k$. If $k$ is odd, then the only
real root $x_{0}=-1+\frac{1}{\sqrt[k]{2}}$ belongs to $\left(  -1,0\right)  $.
If $k$ is even, then the only real roots $x_{1,2}=-1\pm\frac{1}{\sqrt[k]{2}}$,
while $x_{1}\in\left(  -2,-1\right)  $ and $x_{2}\in\left(  -1,0\right)  $.
\end{corollary}

\begin{corollary}
\label{Cor2}The polynomial $I(A_{2k-1};x)$ has only one real root for every
odd $k$, and exactly two real roots for each even $k$. If $k$ is odd, then the
only one real root belongs to $\left(  -1,0\right)  $. If $k$ is even, then
one root belongs to $\left(  -3,-2\right)  $, and the other belongs to
$\left(  -1,0\right)  $.
\end{corollary}

Theorem \ref{th4} claims that
\[
I(\overline{A_{n}};x)=\left(  1+x\right)  \cdot I(A_{n-1};x),n\geq2.
\]
Therefore, the set of roots of $I(\overline{A_{n}};x)$ is the union of the set
of roots of $I(A_{n-1};x)$ and $\{-1\}$.

\section{Conclusions}

It is amusing, but antiregular graphs are, actually, very \textquotedblleft
regular\textquotedblright. One can easily see their pattern, when antiregular
graphs are considered in the context of threshold graphs. Namely, their
building binary strings are of the forms $\mathbf{00}1010101...$ and
$\mathbf{01}010101...$. Let us define a $(a,b)$\textit{-pattern graph} as a
graph with the building binary string of the form $a_{1}a_{2}...a_{q}\left(
b_{1}b_{2}...b_{p}\right)  $, where all $a_{i},b_{j}\in\{0,1\}$, and the
sequence $b_{1}b_{2}...b_{p}$ is a periodic part of the string. For instance,
connected antiregular graphs may be described as $0\left(  01\right)
$-pattern or $\left(  01\right)  $-pattern graphs. This definition opens an
interesting research program. Its main goal is to recognize such patterns
$(a,b)$ that ensure for the independence polynomial of the corresponding
$(a,b)$-pattern graphs to be unimodal, log-concave, or even to have only real roots.


\begin{thebibliography}{99}                                                                                               %


\bibitem {AlMalSchErdos}Y. Alavi, P. J. Malde, A. J. Schwenk, P. Erd\"{o}s,
\emph{The vertex independence sequence of a graph is not constrained},
Congressus Numerantium \textbf{58} (1987) 15-23.

\bibitem {Behzad}M. Behzad, D. M. Chartrand, \emph{No graph is perfect}, Amer.
Math. Monthly \textbf{74} (1967) 962-963.

\bibitem {BSN07}M. Bousquet- {M}\'{e}lou, S. Linusson, E. Nevo, \emph{On the
independence complex of square grids}, Journal of Algebraic Combinatorics
\textbf{27} (2008) 423-450.

\bibitem {BrownDilNow}J. I. Brown, K. Dilcher, R. J. Nowakowski, \emph{Roots
of independence polynomials of well-covered graphs}, Journal of Algebraic
Combinatorics \textbf{11} (2000) 197-210.

\bibitem {CheHaHeLas}G. A. Cheston, E. O. Hare, S. T. Hedetniemi, R. C.
Laskar, \emph{Simplicial graphs}, Congressus Numerantium \textbf{67} (1988) 105-113.

\bibitem {ChvatHam}V. Chvatal, P. L. Hammer, \emph{Set-packing and threshold
graphs}, Res. Report CORR 73- 21, University Waterloo, 1973.

\bibitem {dem}R. W. Deming, \emph{Independence numbers of graphs - an
extension of the K\"{o}nig-Egerv\'{a}ry theorem}, Discrete Mathematics
\textbf{27} (1979) 23-33.

\bibitem {DoPoTi03}K. Dohmen, A. Ponitz, P. Tittmann, \emph{A new two-variable
generalization of the chromatic polynomial}, Discrete Mathematics and
Theoretical Computer Science \textbf{6} (2003) 69-90.

\bibitem {FishSol90}D.C. Fisher, A.E. Solow, \emph{Dependence polynomials},
Discrete Mathematics \textbf{82} (1990) 251--258.

\bibitem {GutHar83}I. Gutman, F. Harary, \emph{Generalizations of the matching
polynomial}, Utilitas Mathematica \textbf{24} (1983) 97-106.

\bibitem {HoLi94}C. Hoede, X. Li, \emph{Clique polynomials and independent set
polynomials of graphs}, Discrete Mathematics \textbf{125} (1994) 219-228.

\bibitem {KeilsonGerber}J. Keilson, H. Gerber, \emph{Some results for discrete
unimodality}, Journal of American Statistical Association \textbf{334} (1971) 386-389.

\bibitem {LeMa02}V. E. Levit, E. Mandrescu, \emph{On well-covered trees with
unimodal independence polynomials}, Congressus Numerantium \textbf{159} (2002) 193-202.

\bibitem {LeMa03a}V. E. Levit, E. Mandrescu, \emph{On unimodality of
independence polynomials of some well-covered trees}, DMTCS 2003 (C. S. Calude
\textit{et al}. eds.), LNCS \textbf{2731}, Springer-Verlag (2003) 237-256.

\bibitem {LeMa03b}V. E. Levit, E. Mandrescu, \emph{A family of well-covered
graphs with unimodal independence polynomials}, Congressus Numerantium
\textbf{165} (2003) 195-207.

\bibitem {LeMa04}V. E. Levit, E. Mandrescu, \emph{Very well-covered graphs
with log-concave independence polynomials}, Carpathian Journal of Mathematics
\textbf{20} (2004) 73-80.

\bibitem {LeMa05}V. E. Levit, E. Mandrescu, \emph{The independence polynomial
of a graph - a survey}, Proceedings of the $1^{st}$ International Conference
on Algebraic Informatics, Aristotle University of Thessaloniki, Greece, 20-23
October, 2005, pp. 233-254.

\bibitem {LeMa06a}V. E. Levit, E. Mandrescu, \emph{Independence polynomials of
well-covered graphs: Generic counterexamples for the unimodality conjecture},
European Journal of Combinatorics \textbf{27} (2006) 931-939.

\bibitem {LeMa06b}V. E. Levit, E. Mandrescu, \emph{Independence polynomials
and the unimodality conjecture for very well-covered, quasi-regularizable, and
perfect graphs}, Graph Theory in Paris: Proceedings of a Conference in Memory
of Claude Berge (A. Bondy et al., eds), Birkh\"{a}user 2006, pp. 243-254.

\bibitem {LeMa06c}V. E. Levit, E. Mandrescu, \emph{Partial unimodality for
independence polynomials of K\"{o}nig-Egerv\'{a}ry graphs}, Congressus
Numerantium \textbf{179} (2006) 109-119.

\bibitem {LeMa07}V. E. Levit, E. Mandrescu, \emph{A family of graphs whose
independence polynomials are both palindromic and unimodal}, Carpathian
Journal of Mathematics \textbf{23} (2007) 108--116.

\bibitem {LeMa08a}V. E. Levit, E. Mandrescu, \emph{Graph operations and
partial unimodality of independence polynomials}, Congressus Numerantium
\textbf{190} (2008) 21-31.

\bibitem {LeMa08b}V. E. Levit, E. Mandrescu, \emph{On the roots of
independence polynomials of almost all very well-covered graphs}, Discrete
Applied Mathematics \textbf{156} (2008) 478 -- 491.

\bibitem {Merris2001}R. Merris, \emph{Graph theory}, Wiley-Interscience,\emph{
}New York, 2001.

\bibitem {Merris2003}R. Merris, \emph{Antiregular graphs are universal for
trees}, Publ. Electrotehn. Fak. Univ. Beograd, Ser. Mat. 14 (2003) 1-3.

\bibitem {Munarini09}E. Munarini, \emph{Characteristic, admittance and
matching polynomials of an antiregular graph,} Applicable Analysis and
Discrete Mathematics \textbf{3} (2009) 157-176.

\bibitem {Nebesky}L. Nebesk\'{y}, \emph{On connected graphs containing exactly
two points of the same degree}, \v{C}asopis P\v{e}st. Mat. \textbf{98} (1973) 305-316.

\bibitem {ProdTichy}H. Prodinger, R. F. Tichy, \emph{Fibonacci numbers of
graphs}, Fibonacci Quart. \textbf{20} (1982) 16-21.

\bibitem {Sedlacek}J. Sedl\'{a}\v{c}ek, \emph{Perfect and quasiperfect
graphs}, \v{C}asopis P\v{e}st. Mat. \textbf{100} (1975) 135-141.

\bibitem {ster}F. Sterboul, \emph{A characterization of the graphs in which
the transversal number equals the matching number}, Journal of Combinatorial
Theory \textbf{B} \textbf{27} (1979) 228-229.

\bibitem {Stevan}D. Stevanovi\'{c}, \emph{Clique polynomials of threshold
graphs}, Univ. Beograd Publ. Elektrotehn. Fac., Ser. Mat. \textbf{8} (1997) 84-87.

\bibitem {Zykov}A. A. Zykov, \emph{Fundamentals of graph theory},
BCS\ Associates, Moscow, 1990.
\end{thebibliography}
\end{document}